# Divergence of neutron microbeams from planar waveguides

S.V. Kozhevnikov[1]*, V.D. Zhaketov[1], T. Keller[2,3], Yu.N. Khaydukov[2,3], F. Ott[4], F. Radu[5]

[1]Frank Laboratory of Neutron Physics, JINR, 141980 Dubna Moscow Region, Russia
*e-mail: kozhevn@nf.jinr.ru
[2]Max Planck Institut für Festkörperforschung, Heisenbergstr. 1, D-70569 Stuttgart, Germany
[3]Max Planck Society Outstation at FRM-II, D-85747 Garching, Germany
[4]Laboratoire Léon Brillouin CEA/CNRS, IRAMIS, Université Paris-Saclay, F-91191 Gif sur Yvette, France
[5]Helmholtz-Zentrum Berlin für Materialien und Energie, Albert-Einstein Straße 15, D-12489 Berlin, Germany

Neutron planar waveguides are focusing devices generating a narrow neutron beam of submicron width. Such a neutron microbeam can be used for the investigation of local microstructures with high spatial resolution. The essential parameter of the microbeam is its angular width. The main contribution to the microbeam angular divergence is Fraunhofer diffraction on a narrow slit. We review and discuss various ways to characterize the angular divergence of the neutron microbeam using time-of-flight and fixed wavelength reflectometers.

## I. INTRODUCTION

Neutron scattering is a powerful tool for the investigation of polymers, biological objects and magnetic systems because of the specific properties of neutrons: high penetration ability, isotopic sensitivity and intrinsic magnetic moment. Neutron scattering is a complementary method to X-rays, for example, for the investigations of magnetic materials in bulk.

Neutron beams at conventional instruments have a width 0.1 - 10 mm. For the investigation of local aperiodic microstructures with high spatial resolution we need very narrow beams. Therefore in the last decades focusing devices in one or two dimensions have been developed, including capillary lenses, Fresnel lenses, elliptical neutron guides and bent crystal monochromators [i]. But these devices cannot achieve microbeam sizes less than 50 μm, restricted by physical properties or technology of the materials used. There are other disadvantages. For example, capillary lenses generate a strong background, Fresnel lenses focus only 20-40 % of the incident beam and elliptical neutron guides produce a beam strongly structured in space and divergence.

More effective focusing devices are planar waveguides. These are tri-layer film where the middle layer with low neutron optical potential is sandwiched by two layers with high neutron optical potential. In [2,3] the polarized neutron microbeam from planar waveguides was used for the investigation of a magnetic microwire with high spatial resolution. The amorphous magnetic wire had axial magnetic domains in a compact core and circular domains in a wide shell [4]. A nonmagnetic waveguide was mounted on neutron reflectometer providing polarization analysis [5]. The method of Larmor precession of neutron spin at transmission [6,7] was used in this experiment. In Fig. 1 the geometry of the experiment is shown. The collimated macrobeam falls onto the waveguide surface under a grazing angle $\delta\alpha_i$, tunnels through an upper thin layer, propagates along the middle layer of width $d \sim 150$ nm and exits from the edge as the microbeam also with width $d$. The microbeam transmits through the investigated sample and is registered by a detector. In this experiment the microbeam was fixed and the sample was translated across the beam.

The spatial resolution is defined by the width $a$ of the microbeam at the sample position at the distance $l \gg d$ and the angular divergence of the microbeam $\delta\alpha_f$ as $a \approx l \cdot \delta\alpha_f$. The main contribution to the angular divergence of the microbeam is Fraunhofer diffraction of the neutron wave on the narrow slit of the width $d$, corresponding to the exit of the waveguide layer:

$$\delta\alpha_F \sim \lambda / d \qquad (1)$$

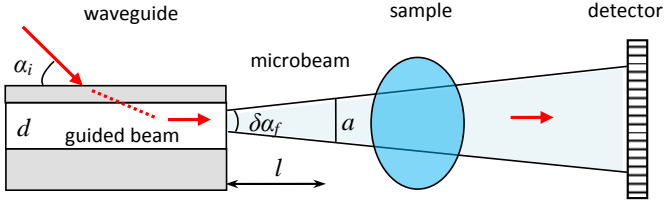

Figure 1: Geometry of an experiment with a neutron microbeam. A planar waveguide produces a very narrow and slightly divergent neutron microbeam emitted from the edge. The investigated microstructure (sample) is scanned across microbeam which is registered by detector. The angular divergence of the microbeam is mainly defined by Fraunhofer diffraction at the waveguide exit. The initial width of the microbeam is equal to $d$ and the width $a$ at the sample position depends on the angular divergence $\delta\alpha_f$ and the distance $l$.

Other contributions to the angular divergence of the microbeam registered by the detector are the angular divergence $\delta\alpha_i$ of the incident beam, the neutron wavelength resolution $\delta\lambda/\lambda$ and the angular resolution of the detector $\delta\alpha_{\det}$. If the neutron wavelength resolution is small in comparison to the angular resolution then we can extract the Fraunhofer diffraction contribution from the experimental angular width of the microbeam:

$$\delta\alpha_F = \sqrt{\left(\delta\alpha_f\right)^2 - \left(\delta\alpha_i\right)^2 - \left(\delta\alpha_{\det}\right)^2} \quad (2)$$

The angular divergence of the microbeam due to Fraunhofer diffraction (1) is directly proportional to the neutron wavelength $\lambda$ and inversely proportional to the guiding layer thickness $d$. In the experiment [2] the estimated width of the neutron microbeam for the distance $l = 1$ mm, the thickness of the guiding layer $d = 150$ nm and the neutron wavelength $\lambda = 4.0$ Å was 2.6 μm. At the time-of-flight reflectometer REMUR [8] a neutron microbeam was registered [9] and the dependence $\delta\alpha_F \sim \lambda$ was measured experimentally. In [10] this dependence was obtained for a set of the waveguides with the different thickness $d$.

In the present work we measured the contribution of Fraunhofer diffraction to the angular divergence of the microbeam at a fixed wavelength reflectometer $\delta\alpha_F \sim 1/d$ and compared the results with time-of-flight data.

## II. NETRON PLANAR WAVEGUIGES

In this section, we will briefly review the theory of neutron resonances in planar waveguides [11]. In Fig. 2 the geometry of planar waveguide is shown schematically. The neutron beam in air (medium 0) enters on the waveguide surface under the grazing angle $\alpha_i$. In Fig. 3a the neutron scattering length density (SLD) is shown as a function of the coordinate $z$ perpendicular to the layers of the waveguide $Ni_{67}Cu_{33}(20$ nm$)/Cu(150)/Ni_{67}Cu_{33}(50)//Si$(substrate). The material $Ni(67$ at.%$)Cu(33$ at.%$)$ is nonmagnetic at room temperature and has a large SLD. The middle layer Cu has a small SLD. Thus, SLD of the waveguide has a shape of a potential well. The neutron wave tunnels through the upper thin layer (medium 1) into the middle layer (medium 2) and is reflected almost totally from the bottom thick layer (medium 3). Then the neutron wave is partially reflected from the upper layer. The component of the neutron wavefunction depending on the coordinate z has a following form:

$$\psi(z) = A\left[\exp\left(-ik_{2z}z\right) + R_{23}\exp\left(ik_{2z}z\right)\right] \quad (3)$$

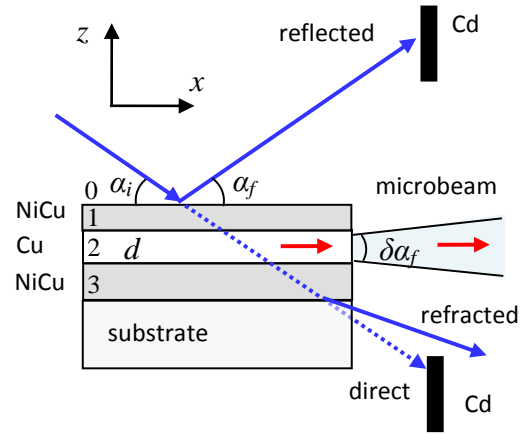

Figure 2: Geometry of planar waveguides.

where $A$ is the amplitude, $k_{2z} = \sqrt{k_{0z}^2 - \rho_2}$ is the $z$-component of the neutron wave vector inside the middle Cu layer, $k_{0z} = k_0 \sin\alpha_i$ is the $z$-component of the neutron wave vector of the incident beam, $\rho_2$ is SLD of the middle Cu layer and $R_{23}$ is the neutron reflection amplitude from the bottom layer. The neutron wavefunction density is resonantly enhanced due to the multiple reflections. The parameter $A$ is determined from the self-consistent equation:

$$A = T_{02} \exp(ik_2d) + R_{21}R_{23} \exp(2ik_2d)A \quad (4)$$

where $T_{02}$ is the neutron transmission amplitude from air (medium 0) through the upper layer into the middle layer (medium 2), $R_{21}$ is the reflection amplitude from the upper layer and $d$ is the thickness of the middle layer. Thus, from (4) we can define:

$$|A| = |T_{02}| / |1 - R_{21}R_{23} \exp(2ik_{2z}d)| \quad (5)$$

The value $k_{0z}$ satisfies the conditions of the resonances when $A$ has maxima:

$$\Phi(k_{0z}) = 2k_{2z}d + \arg(R_{21}) + \arg(R_{23}) = 2\pi n \quad (6)$$

where n=0, 1, 2 ... is the order of resonance, $k_{0z} = \dfrac{2\pi \sin \alpha_{in}}{\lambda}$ for the fixed neutron wavelength mode and $k_{0z} = \dfrac{2\pi \sin \alpha_i}{\lambda_n}$ for the time-of-flight mode.

According to conventional definition, the case $|\psi|^2 \leq 4$ corresponds to the neutron standing waves and such a tri-layer system is termed as *waveguide*. In the case of the resonantly enhanced standing waves at $|\psi|^2 > 4$, this tri-layer structure is termed as *resonator*. In practice, it is difficult to find a tri-layer system without resonant enhancement and the coefficient of the resonant enhancement depends on the quality of the structure (dispersion of layers thicknesses, interface roughness, etc.). The imperfections of the resonator might reduce the resonant enhancement coefficient to magnitude less than 4. Thus, we propose to use the term 'resonator' when the resonant properties of the tri-layer structure are used. When the propagation of the neutron wave along the middle layer is used, we propose to use the term 'waveguide'.

In Fig. 3b the neutron wavefunction density calculated for the fixed neutron wavelength 4.26 Å is shown as a function of the coordinate $z$ perpendicular to the layers and the grazing angle of the incident beam. The dashed line corresponds to the critical angle for total reflection. One can see the maxima for the resonances n = 0, 1, 2, 3. In Fig. 3c, the neutron wavefunction density is plotted vs. $z$. The curves 1, 2 and 3 correspond to the resonances n = 0, 1, 2, respectively.

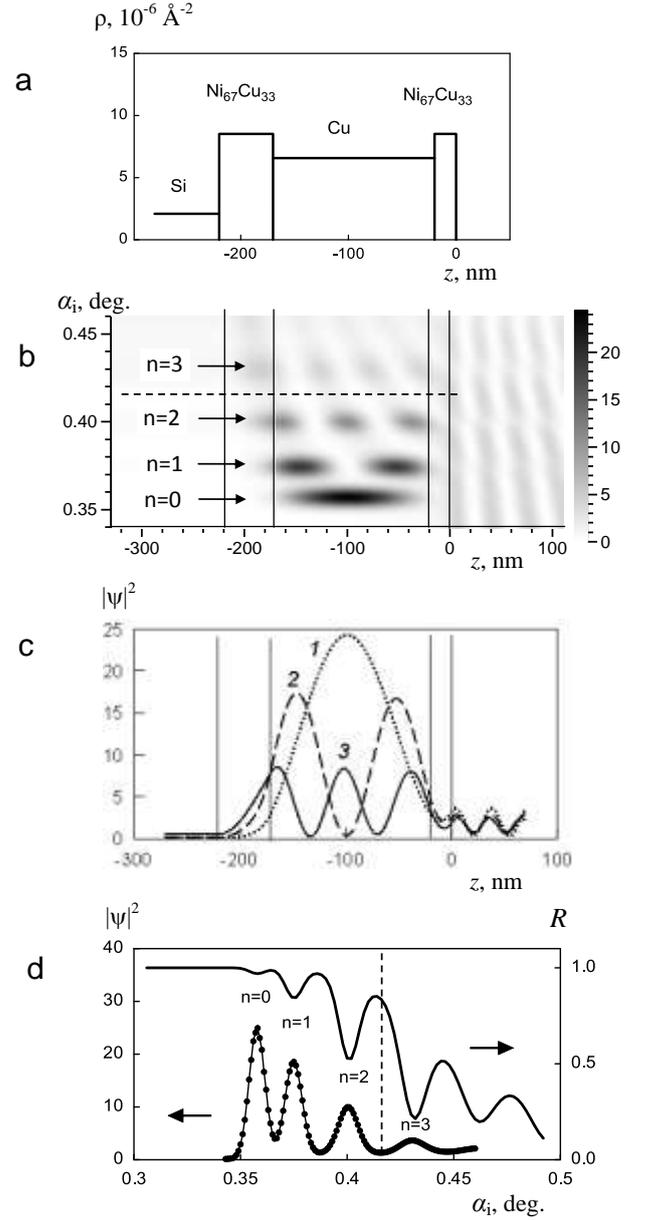

Figure 3: Calculations for the waveguide $Ni_{67}Cu_{33}(20 \text{ nm})/Cu(150)/Ni_{67}Cu_{33}(50)//Si(\text{substrate})$. (a) SLD as a function of the coordinate z perpendicular to the layers. (b) Neutron wavefunction density vs. z and the grazing angle of the incident beam $\alpha_i$ (dashed line is the critical angle of total reflection). The neutron wavelength is 4.26 Å. (c) The neutron wavefunction density vs. z (curve 1, 2 and 3 correspond to the resonances n= 0, 1 and 2, respectively). (d) The neutron wavefunction density (points and left axis) and the reflectivity (right axis) vs. the grazing angle of the incident beam, the dashed line marks a critical angle for total reflection.

One can see one, two and tree maxima for the resonances n = 0, 1, 2, respectively. In Fig. 3d the neutron wavefunction density (points and left axis) and the reflectivity (line and right axis) are shown as a function of the grazing angle of the incident beam. The dashed line corresponds to the critical angle of total reflection. The maxima of the neutron wavefunction density correspond to the weak minima of the reflectivity. To see deep resonances at total reflection it is necessary to have a perfect structure, a very well collimated incident neutron beam and strong absorption in the resonator.

The resonantly enhanced neutron wave is propagating along the guiding layer like in a channel (Fig. 2). Therefore this phenomenon is termed *channeling*. During the process of channeling the neutrons partly leak from the channel through the upper layer and exit in the direction of the specularly reflected beam. This channeled and reflected beam has a resonant nature but has the same width as the specularly reflected macrobeam in contrast to the narrow microbeam from the edge. Due to the leakage of neutrons from the channel through the upper layer the neutron wavefunction density decays exponentially in the direction along the channel as $\sim \exp\left(-x/x_e\right)$ where $x_e$ is the channeling length. According to the theory of neutron channeling in planar waveguides [12] the neutron channeling length depends on the parameters of the structure (upper layer thickness, channel width, potential well depth), the order of the resonances n = 0, 1, 2, ... and imperfections of the structure (dispersion of the layers thickness, interface roughness, etc.). The neutron channeling length in planar waveguides was measured experimentally for the first time in [13,14]. In [15] we describe and discuss the channeling length measurement using an absorber on top of the waveguide surface. The channeling length was measured in dependence on the upper layer thickness and the resonance order [16], the potential depth [17], and the channel width [18]. For example, in [16,19] the measured channeling length for the waveguide in Fig. 3a was $x_e$=(1.7±0.2) mm.

There are several types of tri-layer systems with very similar potential well structure as in Fig. 3a but for different usage. To clarify this point, we briefly review literature. If the upper, middle and bottom layers are thin, then the transmitted beam has one strong resonance like in Fig. 3d and a corresponding deep minimum at total reflection. Thus, transmitted neutrons have a very narrow resonance band. Such structures are termed *interference filters*. The SLD of the layer material corresponds to the energy of ultracold neutrons of about 100 neV. Therefore interference filters were used for spectrometry of ultracold neutrons. The first multilayer interference filter was designed in 1974 [20]. The first experiments with neutron interference filters are discussed in [21,22]. The applications of interference filters in the fundamental experiments with ultracold neutrons are reviewed in [23].

If the upper and bottom layers are thin but the middle layer is relatively thick as in Fig. 3a then we observe many resonances in the region of total reflection. In transmission there are strong resonant maxima and in reflection there are corresponding deep resonant minima. This phenomenon is termed *frustrated total reflection*. Such tri-layer structures are the neutron analog of *Fabry-Perot interferometer*. In 1977 [24] the neutron analog of Fabry-Perot interferometer based on tri-layer structures was proposed. Ref. [25] reviews the experiments with these devices which are very sensitive to the parameters of the layered structure.

Resonators in Fig. 3a are used for the enhancement of neutron interaction with matter which can be registered by several ways: 1) dips at total neutron reflection (primary neutron channel); 2) maxima of neutron intensity (secondary neutron channel), namely off-specular neutron scattering on interface roughness, incoherent neutron scattering at interaction with hydrogen, spin-flipped neutrons at interaction with magnetically non-collinear layers, neutron channeling; 3) maxima of secondary irradiation including gamma-rays, alpha-particles, protons, tritons, and fission products.

For the first time in 1994 the layered resonator was used for observation of resonantly enhanced neutron interaction with matter: in [26] the dips on total neutron reflection from a layered polymer film were observed and in [27] the dips on total neutron reflection and maxima of gamma-irradiation were observed for the layer $Gd_2O_3$. In [28] the resonant minima at total neutron reflection and resonant maxima of alpha-particles intensity were registered for a layer of $^6LiF$. The spin-flipped neutron intensity in specular and off-specular regions was observed for a thin magnetic Co layer placed inside a resonator [29]. The enhanced off-specular neutron scattering was observed in [30,31] due to interface roughness. Additionally, in [30] the resonant maxima of spin-flip neutron intensity were observed when upper and bottom layers were magnetic. In [32-34] off-specular scattering of polarized neutrons was observed for the domain structure near interfaces. The review of methods of registration and application of neutron standing waves in layered structures was done in [35].

Recently, the interest to use layered resonators is increasing again. For example, in [36] a layered resonator was used for the investigations of coexistence of magnetism and superconductivity. In [37,38] the polarized neutron beam was used to change the potential well structure and thus select the different layers for enhanced neutron interaction. In [39] a magnetic layered resonator with uranium inside was proposed to create a compact atomic electrical power station. Using an applied magnetic field it is possible to change the magnetization of the external magnetic layers and the potential well depth. This changes the coefficient of enhancement of the neutron wavefunction density inside the resonator with uranium leading to reactivity modulation for the fission reaction. In [38] the resonant maxima of incoherent neutron scattering from hydrogen containing layers were registered directly.

A planar waveguide to produce a neutron microbeam was considered theoretically for the first time in 1973 in [40]. Such a waveguide is called 'prism-like waveguide' and consists of two parts. The first part is the resonant beam coupler with a thin upper layer. The second part is the waveguide with thick upper layer. For the first time the neutron microbeam from the exit face of the prism-like waveguide was observed in 1998 in [41]. Neutron channeling was observed for the first time in 1994 [42] in the prism-like waveguide in reflection geometry. In this case the waveguide had three parts: resonant beam-coupler, waveguide and resonant decoupler (the same as the first part). The prism-like waveguides have a rather complicated structure and therefore were not used broadly. The simple waveguide based on tri-layer structures as in Figs. 2 and 3a is a more simple and effective device. The neutron channeling in the simple waveguide was observed for the first time in [43] in reflection geometry. Unpolarized [44,45] and polarized [46] neutron microbeams were obtained from the edge of the simple waveguide. The polarizing magnetic waveguide Fe/Co/Fe was investigated in [47]. The polarized neutron channeling can be used for the direct determination of the magnetization value of weakly magnetic films with high accuracy. Such materials containing rare-earth elements are promising for magnetic recording and switching [48]. In [49] the polarized neutron channeling method was proposed and calculations were done for the prism-like waveguide. In [50] calculations were made for the simple waveguides. In [51] this method of polarized neutron channeling was demonstrated experimentally for the film TbCo$_5$. The sensitivity of the direct determination of magnetization value is about 10 G. In [52-54] different neutron methods for the investigation of magnetic films are discussed: Larmor spin precession, Zeeman spatial beam-splitting, neutron spin resonance in matter, and polarized neutron channeling.

## III. FRAUNHOFER DIFFRACTION

The angular distribution of the normalized microbeam intensity is defined by Fraunhofer diffraction as:

$$I = \left( \sin \beta \, / \, \beta \right)^2 \qquad (7)$$

$$\beta = 1/2 \, k_0 d \sin \alpha_F \approx \pi d \alpha_F \, / \, \lambda \qquad (8)$$

where $k_0 = 2\pi \, / \, \lambda$, $\alpha_F \ll 1$ and $\sin \alpha_F \approx \alpha_F$. In Fig. 4a the calculations of the angular distribution of the microbeam intensity for the neutron wavelength $\lambda = 4.26$ Å and the waveguide channel width $d = 141.7$ nm using eqs. (7) and (8) are shown. The angular width (FWHM) of the microbeam $\delta \alpha_F$ is defined from Fig.4a as a graphical solution of the equation

$$\left( \sin \beta \, / \, \beta \right)^2 = 0.5 \qquad (9)$$

As $\beta =$ const, from (8) it follows that $\delta \alpha_F \sim \lambda \, / \, d$.

In Fig. 4b the angular width $\delta \alpha_F$ (see Fig. 4a) was calculated as a function of the neutron wavelength for the fixed $d$ : 80 nm (curve 1), 100 nm (curve 2), 141.7 nm (curve 3) and 180 nm (curve 4). The angular divergence of the microbeam is linearly increasing with the neutron wavelength increasing. The coefficient of proportionality decreases with increasing width of the channel. In Fig. 4c the angular width $\delta \alpha_F$ is calculated as a function of the channel width $d$ for the fixed neutron wavelength 6 Å (curve 1), 4.26 Å (curve 2) and 2 Å (curve 3). The dependence is $\delta \alpha_F \sim 1 / d$ and the coefficient of proportionality is increased by increasing of the neutron wavelength. In Fig. 4d the angular width $\delta \alpha_F$ is presented as a function of $1 / d$ for the fixed neutron wavelength 6 Å (curve 1), 4.26 Å (curve 2) and 2 Å (curve 3). The dependence is linear and the coefficient of the direct proportionality is increasing with the neutron wavelength increasing.

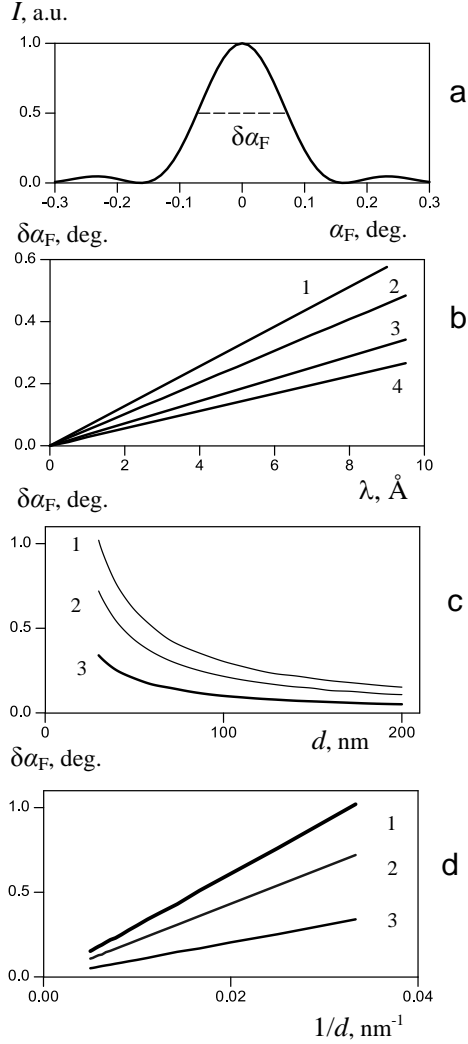

Figure 4: Calculations for Fraunhofer diffraction on a narrow slit of the width $d$. (a) Normalized neutron intensity vs. final angle $\alpha_F$ calculated for the neutron wavelength 4.26 Å and the slit width 141.7 nm. The angular width of the neutron microbeam $\delta\alpha_F$ is defined as FWHM of the central peak. (b) The angular width of the microbeam vs. neutron wavelength for the fixed width $d$ : 80 nm (curve 1), 100 nm (curve 2), 141.7 nm (curve 3) and 180 nm (curve 4). (c) The angular width of the microbeam vs. the channel width $d$ for the fixed neutron wavelength: 6 Å (curve 1), 4.26 Å (curve 2) and 2 Å (curve 3). (d) The angular width of the microbeam as a function of $1/d$ for the fixed neutron wavelength: 6 Å (curve 1), 4.26 Å (curve 2) and 2 Å (curve 3).

Eqs. (7), (8) correspond to Fraunhofer diffraction on a narrow slit with a homogeneous distribution of a neutron wavefunction density across the slit. For the resonance $n = 0$ this distribution is close to the homogeneous one. But for the resonances $n > 0$ the spatial distribution of the neutron wavefunction density inside the waveguide strongly depends on the coordinate $z$. In this case it is necessary to use the Fourier transform of the neutron wave function density for the calculation of the angular distribution of the neutron intensity [44]:

$$I_n\left(\alpha_f\right) = B\left|\int_{-\infty}^{+\infty} \Psi_n\left(z\right)e^{ik_0\sin(\alpha_f)z}dz\right|^2 \qquad (10)$$

where $B$ is the normalization factor which is defined from the fit. In eq. (10) we should include the angular resolution of the incident beam, the angular resolution of the detector and the neutron wavelength resolution.

## IV. EXPERIMENT

### A. Fixed wavelength mode

In this section we present the experimental results obtained on the fixed wavelength reflectometer NREX (reactor FRM II, MLZ, Garching, Germany) with horizontal sample plane. The spatial resolution of the $^3$He two-dimensional position-sensitive detector (PSD) was 3 mm. The distance from the first slit after a monochromator to the sample was equal to 2200 mm and from the sample to the detector was 2400 mm. The second slit of the width about 0.7 mm was placed at the distance 200 mm before the sample and served for background suppression. Two Cd beam-stops were used for the reduction of the specularly reflected and the direct beams (Fig. 2).

Two series of experiments were carried out. In the first experiment, the waveguide with nominal structure Ni$_{67}$Cu$_{33}$(20 nm)/Cu(150)/Ni$_{67}$Cu$_{33}$(50)//Si(substrate) was investigated (Fig. 3). The substrate sizes were $25\times25\times1$ mm$^3$. The first slit height was 0.35 mm. The angular divergence of the incident beam was 0.009° and the neutron wavelength resolution was 2 % (FWHM). The neutron intensity as a function of the grazing angles of the incident beam $\alpha_i$ and the scattered beam $\alpha_f$ is shown in Fig. 5a. The upper diagonal is the part of the specularly reflected beam. The bottom diagonal is the part of the direct beam. For the angles $\alpha_i > 0.3°$ the direct and refracted beams are outside the detector window and are not registered by PSD. The horizontal line $\alpha_f = 0$ corresponds to the sample plane direction. The spots marked by the ellipses are the microbeams of the resonances n = 0, 1, 2. For the resonance n=2 the central spot near the sample plane is not visible because of the weak intensity close to background level. But this weak maximum is visible in the slice vs. $\alpha_f$. In Fig. 5b the reflectivity is shown as a function of the grazing angle of the incident beam $\alpha_i$ (left scale). The right scale corresponds to the specularly reflected beam intensity. The points are the experimental data and the line is fit. The parameters of the structure obtained from fit are following:

Ni$_{67}$Cu$_{33}$O(1.1 nm, 3.31·10$^{-6}$ Å$^{-2}$)/
Ni$_{67}$Cu$_{33}$(18.5, 8.73·10$^{-6}$)/Cu(141.7, 6.58·10$^{-6}$)/
Ni$_{67}$Cu$_{33}$(47, 8.53·10$^{-6}$)// Si(2.07·10$^{-6}$)

In Fig. 5c the neutron microbeam intensity is shown as a function of the grazing angle of the incident beam $\alpha_i$ integrated over the final angle $\alpha_f$ between the reflected and refracted beams. There are the maxima corresponding to the resonances n = 0, 1, 2. The neutron microbeam intensity as a function of the grazing angle of the scattered beam $\alpha_f$ at the fixed angles $\alpha_{in}$ for the resonances n = 0, 1, 2 is presented in Fig. 5d. The points are experimental data and lines are fit using Fourier transformation (10) of the neutron wavefunction density. The angular divergence of the incident beam, the angular resolution of PSD and the neutron wavelength resolution were included in calculations. Calculations describe the experimental data very well. One can see the weak central maximum for the resonance n=2 in contrast to Fig. 3c where the central peak for the neutron wavefunction density is strong. This fact corresponds to Fourier transformation (10) and experimentally confirmed in [44].

The second set of the experiments at NREX reflectometer was carried out for the structures Ni$_{67}$Cu$_{33}$(20 nm)/Cu($d$)/Ni$_{67}$Cu$_{33}$(50)//Al$_2$O$_3$ (substrate) where $d$ = 80, 100, 120 and 180 nm (Fig. 6a). The samples with $d$ = 80, 120 and 180 nm had the sizes 10×10×0.5 mm$^3$ and the sample with $d$ = 100 nm had the sizes 10(along the beam)×20×1 mm$^3$. The first slit height after the monochromator was equal to 0.25 mm, corresponds to the angular divergence of the incident beam of 0.0065°. The neutron wavelength resolution was 1 % (FWHM). The PSD angular resolution was the same as before. For the Al$_2$O$_3$ substrate the value of SLD is close to SLD for Cu in contrast to the case of Si substrate (compare with Fig. 3a).

Therefore the refracted beam is close to the microbeam for the resonance n = 0. In Fig. 6b neutron intensity for the sample with the channel width $d$ = 180 nm is shown as a function of the grazing angles of the incident beam $\alpha_i$ and the scattered beam $\alpha_f$. In the bottom, the neutron beam refracted in the Al$_2$O$_3$ substrate is seen. The neutron microbeams of the resonances n = 0, 1, 2, 3, 4 are marked by the ellipses. One can see that the bottom part of the microbeams is overlaped by the refracted beam.

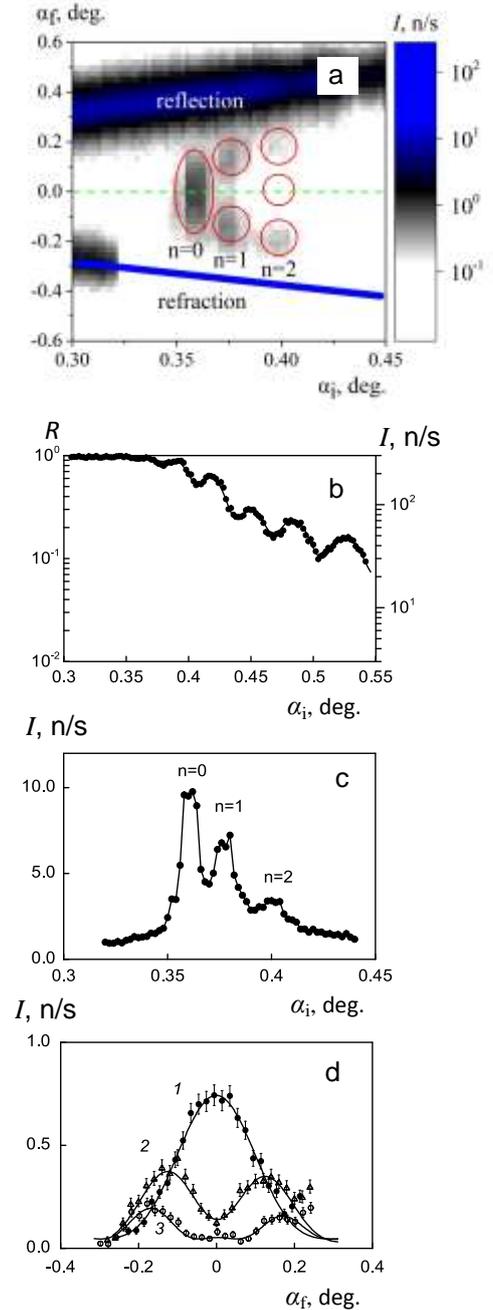

Figure 5: Results for the waveguide Ni$_{67}$Cu$_{33}$(20 nm)/Cu(150)/Ni$_{67}$Cu$_{33}$(50)//Si(substrate). (a) Two-dimensional map of intensity. (b) Reflectivity as a function of the grazing angle of the incident beam (left scale is reflectivity, right scale is intensity). (c) Neutron microbeam integrated intensity as a function of the grazing angle of the incident beam. (d) Neutron microbeam intensity as a function of the grazing angle of the scattered beam at the fixed angles $\alpha_{in}$: n=0 (curve 1), n=1 (curve 2) and n=2 (curve 3). Points are experimental data and lines are fit.

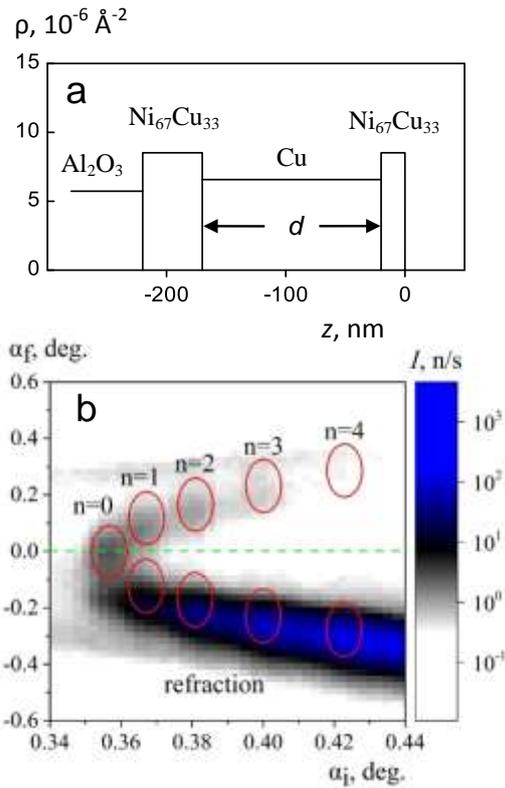

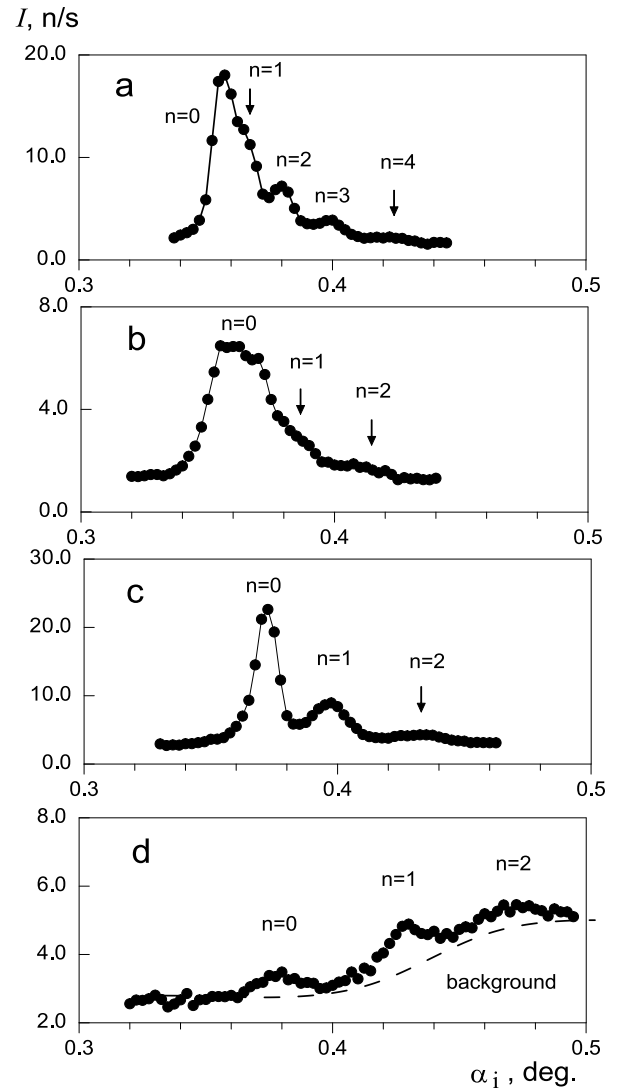

Figure 6: (a) SLD of the waveguide structure $Ni_{67}Cu_{33}(20 \text{ nm})/Cu(d)/Ni_{67}Cu_{33}(50)//Al_2O_3$ (substrate) as a function of the coordinate $z$ perpenicular to the surface. (b) Neutron intensity map for the waveguide with the channel width $d$ = 180 nm as a function of the grazing angle of the incident beam. Ellipses mark the microbeams of the resonances n = 0, 1, 2, 3, 4.

Figure 7: The neutron microbeam intensity integrated between the refracted and the reflected beams as a function of the glancing angle of the incident beam for the samples with different width of the channel $d$ : (a) 180 nm; (b) 120 nm; (c) 100 nm; (d) 80 nm. The indices n = 0, 1, 2, 3, 4 mark the resonances maxima.

In Fig. 7 the neutron microbeam intensity integrated between the refracted and reflected beams for the samples with the channel width 180 nm (a), 120 nm (b), 100 nm (c) and 80 nm (d) is shown as a function of the glancing angle of the incident beam. The distance between the resonances increases with the width of the channel. For the sample with $d$ = 80 nm background is high. It deals with the problem of a thin substrate which is curved and changes the angle of the specularly reflected beam.

In Fig. 8 the neutron microbeam intensity is presented as a function of the final grazing angle for the different channel thickness and resonance orders: (a) 180 nm, n = 0; (b) 120 nm, n = 0; (c) 100 nm, n = 0; (d) 80 nm, n = 0; (e) 180 nm, n = 1; (f) 180 nm, n = 2; (g) 180 nm, n = 3; (h) 180 nm, n = 4. From the left side one can see the refracted beam and from the right side the part of the reflected beam is seen. Points are experimental data and bold line is the calculations using Fourier transformation (10). One can see that the calculations are satisfactory describe the experimental data. For the microbeams of the resonance n = 0 in Figs. 5d and 8a-c we can experimentally define the angular width of the peaks. For this, we describe the shape of the peaks using the function of Gauss which is close to Fourier transformation for the resonance n = 0. But for the Gaussian we do not use the parameters of the waveguides. The width of the Gaussian gives us the experimental value $\delta\alpha_f$ of the angular width of the microbeam n = 0. To obtain the Fraunhofer diffraction contribution $\delta\alpha_F$ from the experimental value $\delta\alpha_f$ we have to extract the angular divergence of the incident beam and PSD angular resolution using eq. (2).

In Fig. 9a Fraunhofer diffraction contribution to the neutron microbeam angular width of the resonance order n = 0 is shown as a function of the channel $d$. Points are experimental data corrected on the angular resolution of the incident beam and PSD using (2). The error bar is defined by the statistics of the neutron count in Figs. 5d and 8a-c. In Fig. 9b the same is presented as a function of $1/d$. One can see that the experimental data are described by Fraunhofer diffraction on a narrow slit.

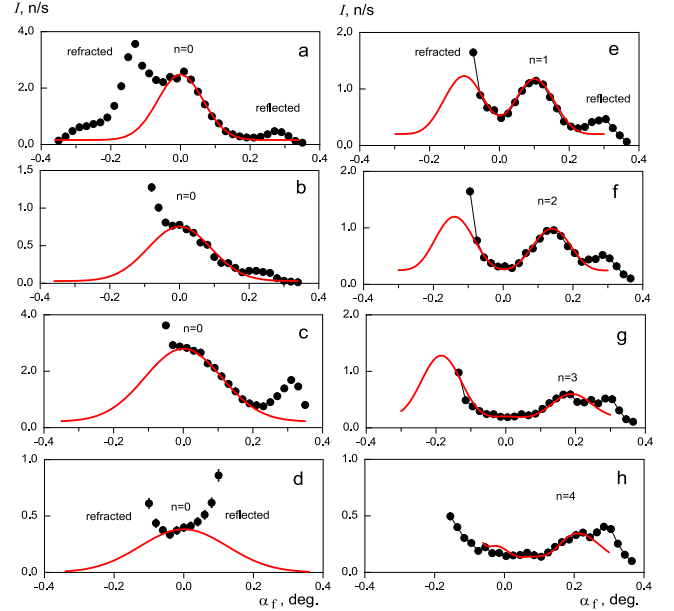

Figure 8: The neutron microbeam intensity as a function of the grazing angle of the reflected beam for the different width of the channel $d$ and the resonance order: (a) 180 nm, n=0; (b) 120 nm, n=0; (c) 100 nm, n=0; (d) 80 nm, n=0; (e) 180 nm, n=1; (f) 180 nm, n=2; (g) 180 nm, n=3; (h) 180 nm, n=4. Points are experimental data, bold line is calculations using (10).

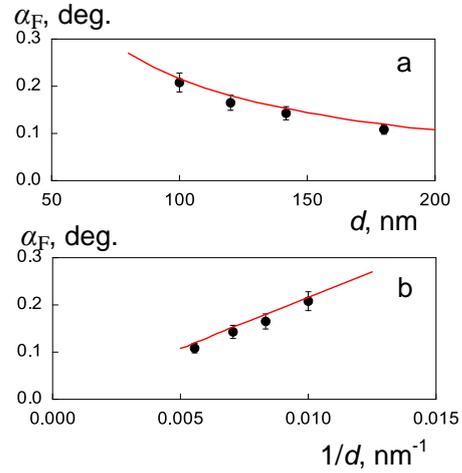

Figure 9: Fraunhofer diffraction contribution into the angular width of the microbeam of the resonance n=0 as a function of: (a) the channel width $d$; (b) the value $1/d$. Points are the experimental values corrected as (2) and line is calculations for the neutron wavelength 4.26 Å using (7)-(9).

## B. Time-of-flight mode

In time-of-flight mode the grazing angle of the incident angle $\alpha_i$ is fixed and the neutron wavelength of the microbeam $\lambda_n$ is defined by the resonances n=0, 1, 2, 3 ... The neutron wavelength of the microbeam is changed by changing the angle $\alpha_i$. The experiment was carried out on the reflectometer REMUR [8] with a vertical sample plane at the pulsed reactor IBR-2. The combined moderator containing the thermal and cryogenic parts was used. The neutron Maxwell spectrum of the polarized incident beam on the exit of the single supermirror polarizer is presented in Fig. 10. This combined moderator reduces in 3 times the neutron intensity in the maximum of the spectrum in the region 1 - 2.5 Å but increases the neutron intensity for the neutron wavelength > 2.5 Å up to 10 times compared to a thermal moderator. The neutron wavelength resolution is defined by the width of the reactor pulse 280 μs and equals to 0.0326 Å for the flight path from the moderator to the detector 34030 mm for REMUR. The distance from the first slit to the sample was 3065 mm and from the sample to the detector was 5030 mm. The first slit for the samples had the width 0.5 mm and the spatial resolution of the ³He two-dimensional position-sensitive detector was 2.5 mm. This corresponds to the angular resolution of the incident beam 0.009° and the angular resolution of PSD 0.028°. The detailed description of the experiments can be found in [9,10].

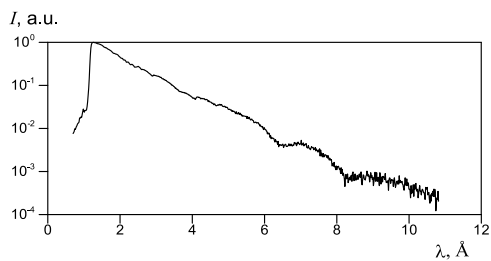

Figure 10: The neutron spectrum of the incident beam at the REMUR time-of-flight reflectometer from the combined (thermal and cryogenic) moderator.

A two-dimensional map of the neutron intensity is shown as a function of the neutron wavelength $\lambda$ and the glancing angle of the scattered beam $\alpha_f$ for the channel width $d = 180$ nm and $\alpha_i = 0.369°$ (Fig. 11a) and $d = 80$ nm and $\alpha_i = 0.246°$ (Fig. 11b). The microbeams are marked by ellipses, the upper and the bottom horizontal lines are the rest of the specularly reflected and direct beams respectively and the bottom curved beam of high intensity is the refracted beam which overlaps the bottom part of the microbeams. For the waveguide with the channel width $d = 180$ nm (Fig. 11a) one can see the microbeams n=0, 1, 2, 3, 4. The microbeam of the resonance n=0 is concentrated near the sample horizon direction $\alpha_f = 0$ and has a small angular divergence. For the sample with the channel width $d = 80$ nm one can see a vertical band corresponding to the large divergent microbeam n=0 near the critical neutron wavelength for refraction in the $Al_2O_3$ substrate.

In Fig. 12 the neutron microbeam intensity is shown as a function of the neutron wavelength $\lambda$ for the different width of the channel 180 nm (a), 120 nm (b), 100 nm (c) and 80 nm (d) and the different glancing angle of the incident beam (curves 1, 2 and 3). In Figs. 12b-d the curve 3 was obtained by summing of the intensity in two neutron wavelength channels for better statistics. In Figs. 12 c,d the curve 3 is multiplied by 10 for clarity. The neutron microbeam intensity is integrated between the reflected and refracted beams. One can see maxima of the neutron intensity corresponding the resonances n=0, 1, 2, 3, 4. The splitting between the maxima positions is increased when the glancing angle of the incident beam is increased and the waveguide channel width is decreased. One can see that the neutron wavelength at the resonance peak position can be increased by increasing the glancing angle of the incident beam.

In Fig. 13 the integrated neutron microbeam intensity for the resonance n=0 is presented as a function of the grazing angle of the scattered beam for the waveguide channel 180 nm (a-c), 120 nm (d-f), 100 nm (g-i) and 80 nm (j-l). The incident angle and the neutron wavelength correspond to the resonances in Fig. 12. Fig. 13a shows the peak of the microbeam n=0 near horizon $\alpha_f = 0$. The neighbor left peak is the refracted beam, the last left peak is the rest of the direct beam and the right peak is the rest of the reflected beam. Bold line is calculations using Fourier transformation (10) with the parameters extracted from the fit of reflectivities [10]. Calculations describe the experimental data within the error bars. The width of the microbeam peak is increased by increasing grazing angle of the incident beam and decreasing of the waveguide channel width. For the channel width 80 nm (Figs. 13j,k,l) broad maxima of the neutron microbeam corresponding to calculations are visible.

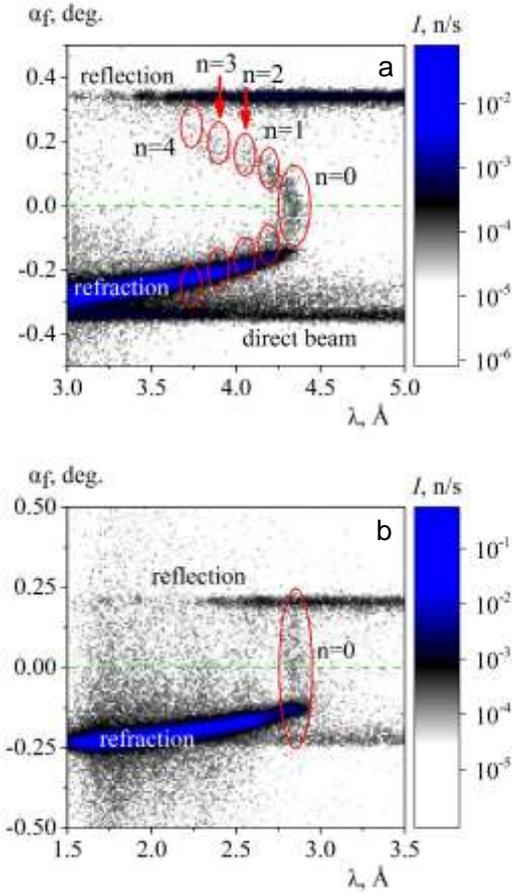

Figure 11: Neutron intensity as a function of the neutron wavelength and the grazing angle of the scattered beam for the different channel width and at the fixed glancing angle of the incident beam: (a) 180 nm, 0.369°; (b) 80 nm, 0.246°.

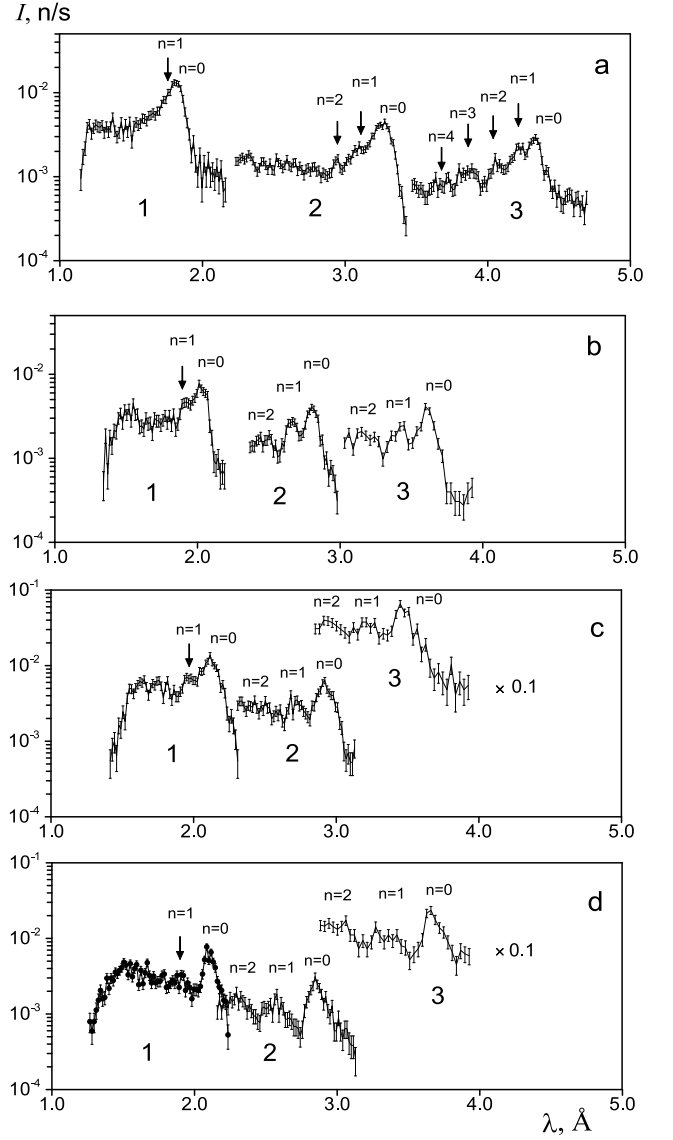

Figure 12: Neutron microbeam intensity as a function of the neutron wavelength for the different waveguide channel width and the glancing angle of the incident beam: (a) 180 nm, 0.152° (curve 1), 0.281° (curve 2), 0.369° (curve 3); (b) 120 nm, 0.169° (1), 0.235° (2), 0.304° (3); (c) 100 nm, 0.175° (1), 0.246° (2); 0.292° (3); (d) 80 nm, 0.177° (1), 0.246° (2), 0.308° (3).

In Fig. 14 the integrated neutron microbeam intensity for the waveguide channel 180 nm is presented as a function of the grazing angle of the scattered beam for the resonances n=0 (a), n=1 (b), n=2 (c), n=3 (d) and n=4 (e). Points are the experimental data and the bold line is calculated using Fourier transformation (10). Calculations describe satisfactory the right part of the neutron intensity distribution. The left part of the microbeams is covered by the refracted beam.

In Fig. 15 the analysis of the experimental angular width of the neutron microbeam n=0 is shown. In Fig. 15a the angular width of the neutron microbeam for the waveguide channel 141.7 nm is presented as a function of the neutron wavelength [9]. Closed symbols correspond to the experimental angular width $\delta\alpha_f$, open symbols correspond to Fraunhofer diffraction contribution $\delta\alpha_F$ extracted from the experimental data using (2) and line is the Fraunhofer diffraction calculations using (7)-(9). One can see that the experimental data for $\delta\alpha_F$ are described by Fraunhofer diffraction calculations.

In Fig. 15b the experimental angular width $\delta\alpha_f$ extracted from the data in Fig. 13 is shown as a function of the neutron wavelength for the waveguide channel width 180 nm (1), 120 nm (2) and 100 nm (3). Points are experimental data and line is linear fit. In Fig. 15c Fraunhofer diffraction contribution $\delta\alpha_F$ extracted from the experimental data using (2) is shown as function of the neutron wavelength for the waveguide channel width 180 nm (curve 1), 141.7 nm (curve 2), 120 nm (curve 3) and 100 nm (curve 4). Points are experimental data and line is a linear fit. The vertical dashed line corresponds to the neutron wavelength 4.26 Å used on the fixed wavelength reflectometer NREX. To change the grazing angle of the incident beam, on the time-of-flight reflectometer REMUR the sample is rotated by a step motor. The minimal step of the motor corresponds to the step of the grazing angle of the incident beam 0.0286°. For the neutron wavelength 4.32 Å and the angle 0.369° (curve 3 in Fig. 12a) the step of the neutron wavelength is equal to 0.34 Å or $\Delta\lambda/\lambda = 7.8\,\%$. In this case we cannot fix the neutron wavelength experimentally for different measurements. But we can use the linear fit in Fig. 15c to extract the experimental Fraunhofer diffraction contribution value $\delta\alpha_F$ at a desirable neutron wavelength. In Fig. 15d Fraunhofer diffraction contribution $\delta\alpha_F$ is shown as a function of the waveguide channel width at the neutron wavelength 4.26 Å (curve 1), 3 Å (curve 2) and 2 Å (curve 3). The points are the experimental value defined from linear fit in Fig. 15c at the fixed wavelength and line is calculations using (7)-(9). One can see that experimental points are well described by calculations. The angular width of the microbeam $\delta\alpha_F$ due to Fraunhofer diffraction is decreasing with increasing the waveguide channel width. In Fig. 15e the angular width of the microbeam $\delta\alpha_F$ due to Fraunhofer diffraction is presented as a function of $1/d$. The linear dependence is obvious. Points are experimental data and the lines show calculations using (7)-(9). The dependence is linear and experiment is described by calculations.

From the experimental data at REMUR we can estimate the angular divergence of the neutron microbeam outgoing from the waveguide edge as $\delta\alpha_f = \sqrt{\left(\delta\alpha_F\right)^2 + \left(\delta\alpha_i\right)^2}$.

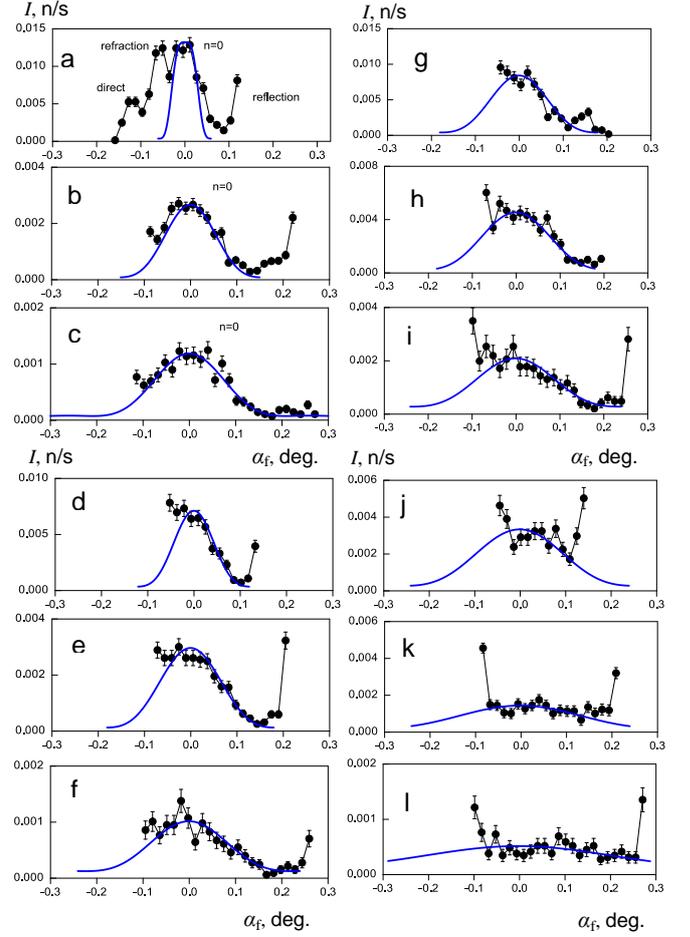

Figure 13: Neutron microbeam intensity of the resonance n=0 as a function of grazing angle of the scattered beam for the different width of the waveguiding channel and the grazing angle of the incident beam: (a) 180 nm, 0.152°; (b) 180 nm, 0.281°; (c) 180 nm, 0.369°; (d) 120 nm, 0.169°; (e) 120 nm, 0.235°; (f) 120 nm, 0.304°; (g) 100 nm, 0.175°; (h) 100 nm, 0.246°; (i) 100 nm, 0.292°; (j) 80 nm, 0.177°; (k) 100 nm, 0.246°; (l) 100 nm, 0.308°.

For the neutron wavelength 4.26 Å, the channel width 180 nm, the angular divergence of the incident beam 0.009° and $\delta\alpha_F = 0.110°$ we can calculate $\delta\alpha_f = 0.110°$. This angular divergence corresponds to the microbeam broadening 1.92 µm/mm. At the distance of 1 mm from the waveguide edge the width of the neutron microbeam is equal to 2.07 µm.

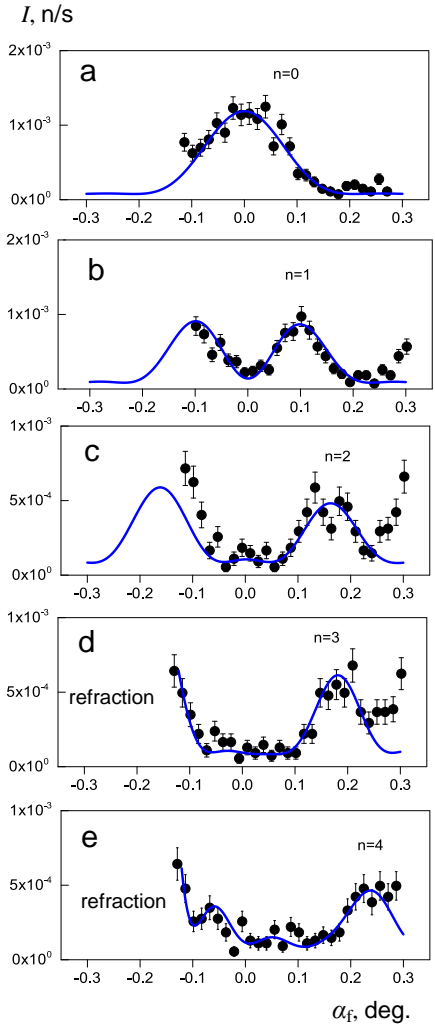

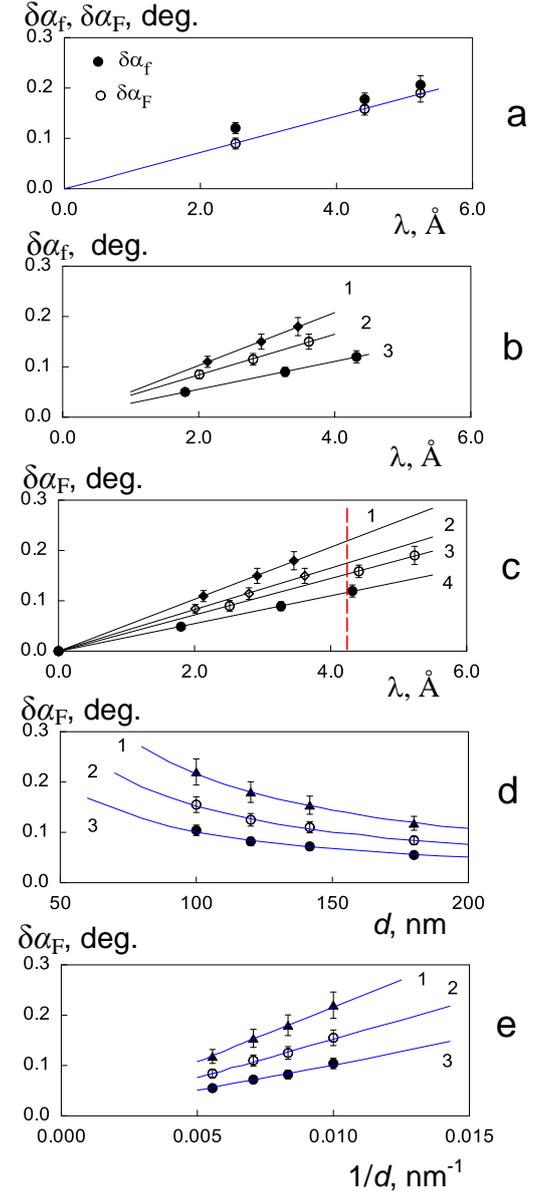

Figure 14: Neutron microbeam intensity of the resonances n=0, 1, 2, 3, 4 as a function of the grazing angle of the scattered beam for the waveguiding channel width 180 nm and the grazing angle of the incident beam 0.369°: (a) n=0; (b) n=1; (c) n=2; (d) n=3; (e) n=4.

Figure 15: (a) The experimental angular width of the neutron microbeam of the resonance n=0 for the channel width 141.7 nm as a function of the neutron wavelength. Closed symbols correspond to the measured width of the peak, open symbols correspond to Fraunhofer diffraction contribution. (b) The experimental angular width of the neutron microbeam of the resonance n=0 for the channel width 100 nm (curve 1), 120 nm (curve 2) and 180 nm (curve 3) as a function of the neutron wavelength. (c) Experimental Fraunhofer diffraction contribution to the angular width of the microbeam n=0 as a function of the neutron wavelength for the channel width 100 nm (curve 1), 120 nm (curve 2), 141.7 nm (curve 3) and 180 nm (curve 4).

Points correspond to experimental data and line is the linear fit. (d) Experimental Fraunhofer diffraction contribution to the angular width of the microbeam n=0 as a function of the waveguiding channel width $d$ for the neutron wavelength 4.26 Å (curve 1), 3 Å (curve 2) and 2 Å (curve 3). Points are experimental data and line is a calculation using (7)-(9). (c) Experimental Fraunhofer diffraction contribution into the angular width of the microbeam n=0 as a function of $1/d$ for the neutron wavelength 4.26 Å (curve 1), 3 Å (curve 2) and 2 Å (curve 3). Points are experimental data and line is calculations using (7)-(9).

## V. DISCUSSION

Neutron sources provide divergent beams with large cross sections and low brilliance. Focusing of such beams on small spots with reflection optics based on the low total reflection angles in the order of one degree is challenging and generally not possible. Modern synchrotron x-ray sources deliver brilliant beams with low cross section and low divergence. Although the total reflection angles comparable to those for neutrons, focusing of x-ray beams to sub-micrometer spots is feasible and frequently used. For neutrons, the use of planar waveguides to generate beam with sub-μm size is an interesting alternative.

Based on the discussion in the preceding sections, a few general rules for the design of planar waveguides generating microbeams can be defined. 1) The width $d$ of the guiding channel should be adapted to the required resolution, but it should not be smaller. 2) A rather short neutron wavelength of about 2 Å is favorable. 3) The investigated sample should be placed as close as possible to the waveguide edge, preferably at a distance of about 1 mm. In this geometry, non-magnetic waveguides and polarization analysis should be used for the investigation of magnetic microstructures, as in this case the use of a magnetic field on the sample does not affect the non-magnetic waveguide. In [2,3] we used electromagnetic coils to rotate the magnetic field with respect to the investigated micro-wire placed at the distance of 1 mm from the waveguide edge. Thus, we have demonstrated the use of an electromagnet in the experiment with the neutron microbeam from the planar waveguide.

The requirement to place the waveguide close to the sample is not a severe limitation, as these waveguides are quite short. As the neutron channeling length is around 2.5 mm, the waveguide length can be limited to about 5 mm, as the gain of longer waveguides is negligible. The short 5 mm waveguides can easily be placed in complicated sample environments, such as cryostats.
In [3] different ways for neutron microbeam shaping are discussed, including narrow slits from absorbing material plates, such as GGG crystals, and total neutron reflection from short substrates, such as Si, under a small grazing angle, and planar waveguides. The most versatile method is total reflection, which has the following advantages: high neutron intensity, compatibility with the time-of-flight technique, a low background level, and a small microbeam broadening in the order of 0.1 - 1 μm/mm. The achievable microbeam width of 30 μm is much larger than those of planar waveguides in the order of 1 μm.

The gain factor of waveguides can be defined as the ratio of the neutron microbeam intensity $I$ over to the neutron intensity of the beam $I_s$ with the same divergence and width formed by hypothetical slits:

$$\eta = I / I_s \qquad (11)$$

This geometry is shown in Fig. 16a. The width of the first slit width is $h_1 = 0.35$ (mm). The intensity of the totally reflected beam is equal to the incident beam intensity. The glancing angle of the incident beam at the resonance $n = 0$ is $0.36°$. In the experiment, the length of the waveguide was 25 mm. The sample selects the incident beam of width $h_2 = 25 \cdot \sin 0.36° = 0.157$ (mm). The intensity of this beam is 290 (n/s), corresponding to the intensity of the totally reflected beam in Fig. 5b for a channel width of 150 nm. In Fig. 16b the equivalent scheme with two slits is shown. The microbeam has the divergence of $0.14°$ and a width of 150 nm. The same divergence (Fig. 16c) can be obtained by setting the first and second slits to $h_1 = 2200 \cdot \sin 0.14° = 5.37$ (mm) and $h_2 = 150$ (nm). The intensity is then

$$I_s = \frac{290 \cdot 5.37 \cdot 0.15 \cdot 10^{-3}}{0.35 \cdot 0.157} = 4.3 \qquad \text{(n/s)}. \qquad \text{The}$$

microbeam intensity in Fig. 5c extracting background of 2.0 n/s is $I = 7.8$ (n/s), corresponding to a gain factor of $\eta = 1.8$.

As the microbeam intensity is about $10^{-2}$ of the reflected and refracted beams, background suppression is crucial. One major contribution for background, especially for a short neutron wavelength, is caused by scattering on the detector window. Therefore beam-stops for the reflected and direct beams should be used. The refracted beam might be blocked by using a substrate of absorbing materials (for example, GGG or boron glass). To separate the refracted beam from the microbeam corresponding the resonance order $n = 0$ at a fixed neutron wavelength, a structure of the waveguide should be adapted by varying the following parameters: the SLD of the substrate, the thickness of the bottom layer with a high SLD, or by choosing a substrate with high SLD, comparable to the SLD of the bottom layer.

To date, we have shown the feasibility of neutron microbeams in several experiments. A polarized neutron microbeam intensity of about 1 n/s at the PRISM reflectometer using a neutron wavelength of 4.0 Å proved sufficient to scan a magnetic microwire in a reasonable measuring time of about 10 hours [2,3].

At the NREX reflectometer, a neutron microbeam intensity 10 n/s for a neutron wavelength 4.26 Å was sufficient to use the polarized neutron channeling method for the investigation of the weakly magnetic films of $TbCo_5$ [16] and $TbCo_{11}$ [55]. Thus we have demonstrated the feasibility of polarized neutron microbeam applications for the investigations of magnetic nanostructures at fixed wavelength reflectometers.

At the time-of-flight reflectometer REMUR, the neutron microbeam intensities of about $10^{-2}$ n/s were achieved, strongly dependent on the neutron wavelength. We used the combined (cryogenic and thermal) moderator to increase the neutron intensity for the large wavelengths. Such a microbeam intensity is sufficient for the investigation of microbeams or waveguides themselves [9,10,56], and also for polarized neutron channeling in the weakly magnetic film $TbCo_{11}$ [57]. This situation is improved at high-flux neutron sources, such as the ESS.

In conclusion, we have investigated the angular width of the neutron microbeam from the edge of the planar waveguides. It was shown experimentally that the main contribution to the angular divergence of the microbeam is Fraunhofer diffraction on a narrow slit of the width $d$ where the guiding layer (or channel) of the waveguide plays the role of the narrow slit. We have measured the angular width of the microbeam and corrected it on the angular divergence of the incident beam and the angular resolution of the position-sensitive detector. At the time-of-flight reflectometer REMUR the dependence $\delta\alpha_F \sim \lambda$ was defined. The dependence $\delta\alpha_F \sim 1/d$ was obtained at the fixed wavelength reflectometer and also at the time-of-flight reflectometer REMUR. Both reflectometers give the same experimental results, which are also well described by calculations.

## Acknowledgements

Authors are thankful to V.K. Ignatovich for fruitful discussions and V.L. Aksenov and Yu.V. Nikitenko for the interest to this subject. This work is supported by JINR-Romania scientific project No. 323/21.05.2018, items 89 and 90. Part of the experimental work was conducted at the NREX reflectometer at the MLZ, Garching. Another part was done at the REMUR reflectometer at FLNP, JINR, Dubna.

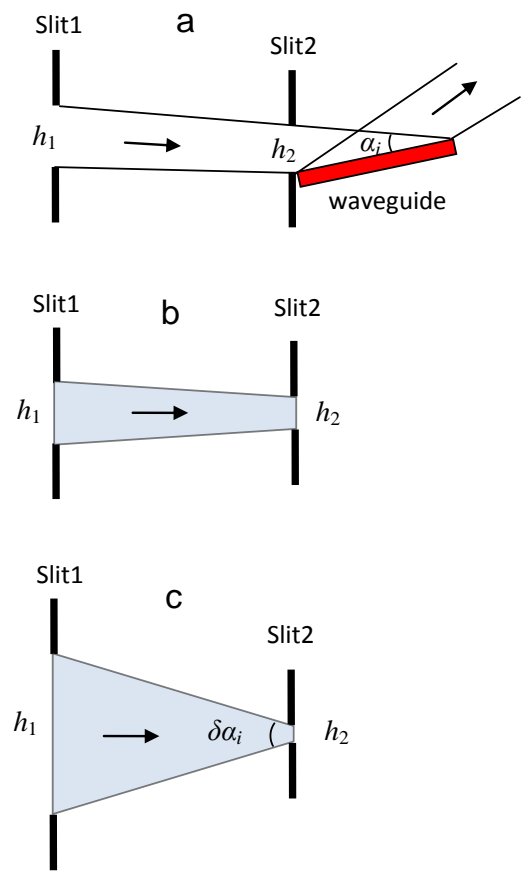

Figure 16: Calculation of the gain factor of waveguides. (a) Total reflection geometry. (b) The equivalent scheme for total reflection. (c) The equivalent scheme for a divergent microbeam.